\documentclass[fleqn,usenatbib]{mnras}

\usepackage[T1]{fontenc}

\usepackage{graphicx}	
\usepackage{amsmath}	
\usepackage{multirow}
\usepackage{subcaption}
\usepackage{amssymb}
\usepackage{tabularx}
\usepackage{hyperref}
\usepackage{txfonts}
\usepackage{makecell}

\usepackage{xcolor}
\usepackage{soul}

\title[Quintom Model Perturbations]{Quintom Model Perturbations}

\author[L.W.K. Goh and A. N. Taylor]{
L. W. K. Goh$^{1,2}$\thanks{E-mail: lgoh@roe.ac.uk}
and A. N.  Taylor$^{1,2}$
\\
$^{1}$Institute for Astronomy, University of Edinburgh, Royal Observatory, Blackford Hill, Edinburgh EH9 3HJ, UK\\
$^{2}$Higgs Centre for Theoretical Physics, School of Physics and Astronomy, The University of Edinburgh, Edinburgh EH9 3FD, UK
}

\date{Accepted XXX. Received YYY; in original form ZZZ}

\pubyear{\the\year{}}

\begin{document}
\label{firstpage}
\pagerange{\pageref{firstpage}--\pageref{lastpage}}
\maketitle

\begin{abstract}
We build upon the work of Goh and Taylor 2025, in which we proposed a two scalar field quintom framework capable of naturally realising a phantom-to-quintessence transition in the dark energy equation of state. In this work, we derive the linear perturbation equations of our model and investigate its implications for large scale structure formation. We show that the quintom model is able to reproduce the phenomenological features of a $w_0w_a$CDM cosmology with phantom-to-quintessence crossing, including suppressions in the matter power spectrum and enhancements to the late-time Integrated Sachs-Wolfe effect. We then perform a Bayesian analysis of the quintom framework with BAO, CMB and Type 1a supernovae data, finding that it is mildly favoured over the standard $w_0w_a$CDM parametrisation, while successfully reproducing physical trends observed in the data. We further examine the parameter degeneracies inherent to the model and discuss prospective observational strategies for distinguishing quintom cosmologies from conventional dynamical dark energy models given current and future data precision.
\end{abstract}

\begin{keywords}
Cosmology:theory -- Dark Energy -- Cosmology:observations
\end{keywords}



\section{Introduction}

The nature of dark energy, responsible for the accelerating expansion of the Universe, remains one of the central open questions in modern cosmology. Within the concordance $\Lambda$CDM model, dark energy is described as a cosmological constant, $\Lambda$, with a constant equation of state $w \equiv p/\rho = -1$. However, recent observations, particularly baryon acoustic oscillation (BAO) measurements from the Dark Energy Spectroscopic Instrument \citep[DESI;][]{2025arXiv250314738D} and Type 1a supernovae \citep[SNe1a;][]{2022ApJ...938..113S,2024ApJ...973L..14D, 2025ApJ...986..231R}, have begun to hint at the possible dynamical nature of dark energy. In such scenarios, the equation of state evolves with time and may transition from $w < -1$ to $w > -1$ at late times when adopting the Chevallier--Polarski--Linder \citep[CPL;][]{2001IJMPD..10..213C,2003PhRvL..90i1301L} parametrisation,
\begin{equation}
    w(a) = w_0 + w_a(1 - a)\,.
\end{equation}
This so-called `phantom crossing' behaviour has traditionally been considered difficult to realise within standard theoretical frameworks, or even scalar field theories in which a scalar field with a canonical kinetic energy term and a specific form of the potential, can act as the dark energy component driving late-time cosmic acceleration \citep{PhysRevD.37.3406,1988NuPhB.302..668W}. 

Such fields with a phantom equation of state (i.e. $w < -1$) typically suffer from ghost instabilities: due to the negative kinetic term in its Lagrangian, the energy is unbounded from below \citep{2002PhLB..545...23C}. Moreover, a single scalar field with a canonical kinetic term cannot cross the phantom divide $w = -1$ by construction, requiring the inclusion of higher-order kinetic terms, as in $k$-essence models \citep{2000PhRvD..62b3511C,2000PhRvL..85.4438A,2001PhRvD..63j3510A}. However, these models generically encounter gradient instabilities \citep{2005PhRvD..71b3515V}, since the effective adiabatic sound speed satisfies $c_{\rm s}^2 \leq 0$ during the crossing. Consequently, additional degrees of freedom are required, capable of realising stable phantom-crossing behaviour. This has motivated the exploration of beyond-$\Lambda$CDM scenarios, for instance subclasses of Horndeski scalar-tensor theories \citep{1974IJTP...10..363H} that modify gravity or introduce couplings to scalar fields \citep[see, e.g.][]{2020PhRvD.101h3507Y, 2024JHEAp..42..217E, 2024JHEP...05..327Y, 2024PhRvD.110l3524C, 2025arXiv250118336Y,2025arXiv250400994P,2025arXiv250320896P,2025arXiv250319898P,2025arXiv250817231T,2025arXiv250314743L,2025arXiv250621010N,2025arXiv250801759W,2025PhRvD.111f1306W,2025arXiv250407679W,2026arXiv260617951B, 2026PhLB..87240096G, 2026arXiv260412032G, 2026arXiv260620794N}.

In \citet{2025MNRAS.544.3142G}, we investigated a quintom model, which postulates two minimally coupled scalar fields acting as dark energy components: one phantom-like and the other quintessence-like. It was first proposed in \citet{2005PhLB..607...35F, 2005PhLB..608..177G} and has also been studied in \citet{2010PhR...493....1C, 2026arXiv260601360R}. This framework allows for a stable realisation of phantom crossing at both the background and perturbative levels. We demonstrated that, by adopting a suitable form of the scalar field potential, specifically one bounded from above by a cut-off, a natural transition from phantom to quintessence behaviour can be achieved. The dynamics of the dark energy sector are governed primarily by the relative evolution rates of the two fields, which are in turn determined by the slope of a common potential.

In this work, we extend our previous analysis by considering the scalar field perturbations and examining their impact on key observables, such as the matter power spectrum and the cosmic microwave background (CMB). With a complete modelling of this two-field system, we modify existing Einstein-Boltzmann solvers and perform a full Bayesian inference analysis of the quintom model. This enables us to determine best-fit parameter values and to assess the model's statistical preference relative to $\Lambda$CDM and $w_0 w_a$CDM cosmologies.

This paper is organised as follows. In Section~\ref{sec:quintom_model}, we review the quintom framework at the background level, focusing on the specific setup introduced in \citet{2025MNRAS.544.3142G}. In Section~\ref{sec:perturb}, we analyse the perturbations and their implications for structure formation, particularly in the context of a phantom-to-quintessence transition, and discuss connections to trends observed in $w_0 w_a$CDM parametrisations. In Section~\ref{sec:mcmc}, we describe the methodology for our Bayesian inference analysis, with results presented in Section~\ref{sec:results}. Finally, we summarise our findings and conclude in Section~\ref{sec:conclusions}.

\section{The Quintom model}\label{sec:quintom_model}

Here, we briefly summarise the relevant background equations of the quintom model, before conducting a more in-depth study of its perturbative behaviour in the following sections. For a detailed analysis of the dynamics of the fields and their consequences on background expansion, we refer the reader to \citet{2025MNRAS.544.3142G} and references therein.

\subsection{Background}

We consider a minimally coupled dark energy sector composed of a quintessence field $\phi$ and a phantom field $\psi$, described by the Lagrangian
\begin{equation}
    \mathcal{L}=X_\phi - X_\psi + V(\phi)+V(\psi)\,,
\end{equation}
where $X_\phi\equiv \tfrac{1}{2}\dot{\phi}^2$ and $X_\psi\equiv \tfrac{1}{2}\dot{\psi}^2$ are the canonical kinetic terms of the respective fields, and $V$ denotes their potentials. The defining feature of a phantom field is its negative kinetic contribution to the Lagrangian, allowing its equation-of-state parameter to satisfy $w<-1$.

Assuming a spatially flat Friedmann--Lemaître--Robertson--Walker (FLRW) background, the Friedmann equation is
\begin{equation}
\label{eq:friedmann}
    H^2=\frac{8\pi}{3}\left(\rho_{\rm m}+\rho_{\rm r}+\rho_\phi+\rho_\psi\right),
\end{equation}
where we work in natural units with $\hbar=c=G=1$, $H\equiv\dot{a}/a$ is the Hubble parameter, and the subscripts `m' and `r' denote non-relativistic matter (baryons and cold dark matter) and radiation respectively. A dot represents differentiation with respect to cosmic time. The total dark energy density is therefore given by the combined contributions of the quintessence and phantom sectors.

The pressures and energy densities of the two fields are
\begin{align}
    p_\phi &= \frac{1}{2}\dot{\phi}^2 - V(\phi)\,, 
    &\rho_\phi &= \frac{1}{2}\dot{\phi}^2 + V(\phi)\,, \label{eq:p_rho_phi}\\
    p_\psi &= -\frac{1}{2}\dot{\psi}^2 - V(\psi)\,, 
    &\rho_\psi &= -\frac{1}{2}\dot{\psi}^2 + V(\psi)\,.
    \label{eq:p_rho_psi}
\end{align}
Each field independently satisfies a Klein--Gordon equation,
\begin{align}
\ddot{\phi}+3H\dot{\phi}+V_{\phi}&= 0\,, 
\label{eq:kg_phi}\\
\ddot{\psi}+3H\dot{\psi}-V_{\psi}&= 0\,,
\label{eq:kg_psi}
\end{align}
where $V_{\phi}=\partial V/\partial \phi$, and the sign difference in the potential derivative term reflects the phantom nature of $\psi$. Consequently, the two fields exhibit opposite dynamical behaviour: as the Universe evolves towards lower redshift, the quintessence field rolls down its potential and its energy density decreases, whereas the phantom field climbs up the potential, causing its energy density to grow with time. For simplicity, we assume that both fields evolve under the same potential form.

Individually, each field remains confined to its respective quintessence or phantom regime. However, the effective equation of state of the combined system,
\begin{equation}
\label{eq:w_tot}
    w_{\rm quintom}=\frac{p_\phi+p_\psi}{\rho_\phi+\rho_\psi}
    =\frac{\frac{1}{2}\dot{\phi}^2-\frac{1}{2}\dot{\psi}^2 -V(\phi)-V(\psi)}
    {\frac{1}{2}\dot{\phi}^2-\frac{1}{2}\dot{\psi}^2 +V(\phi)+V(\psi)}\,,
\end{equation}
can dynamically cross the cosmological-constant boundary $w=-1$, thereby realising the characteristic quintom behaviour. This can be seen particularly clearly in the slow-roll regime, where defining
\begin{equation}
\epsilon \equiv \frac{X}{V}
=
\frac{\dot{\phi}^2-\dot{\psi}^2}
{2\left[V(\phi)+V(\psi)\right]},
\end{equation}
yields
\begin{equation}
w_{\rm quintom}\approx -1+2\epsilon\,.
\end{equation}
The relative rolling speeds of the two fields therefore determine both the timing and direction of the phantom crossing.

Quintom scenarios are commonly classified into two categories: type-A models, in which the Universe evolves from quintessence domination to phantom domination, and type-B models, which exhibit the reverse transition from phantom to quintessence behaviour. Power-law or exponential potentials naturally lead to type-A behaviour, but current observational data mildly favour the latter class. In \citet{2025MNRAS.544.3142G}, we explored a type-B quintom scenario and demonstrated that a hilltop or cliff-face potential can naturally induce a phantom-to-quintessence transition. In particular, we considered a step-like hyperbolic tangent potential of the form
\begin{equation}
\label{eq:v_tanh}
    V(\phi)= V_0\left[\tanh\left(s(1-\phi)\right)+1\right],
\end{equation}
where $V_0$ is the overall amplitude of the potential and $s$ controls the steepness of the cliff face. The introduction of a sharp cut-off near the potential maximum effectively freezes the quintessence field while the phantom field climbs the potential, generating an early phantom dominated phase. As the phantom field reaches the plateau and slows, the quintessence field subsequently rolls down the potential, driving the Universe into a quintessence dominated epoch and naturally producing a phantom-to-quintessence transition.

\begin{figure*} 
    \centering 
    \includegraphics[width=\linewidth]{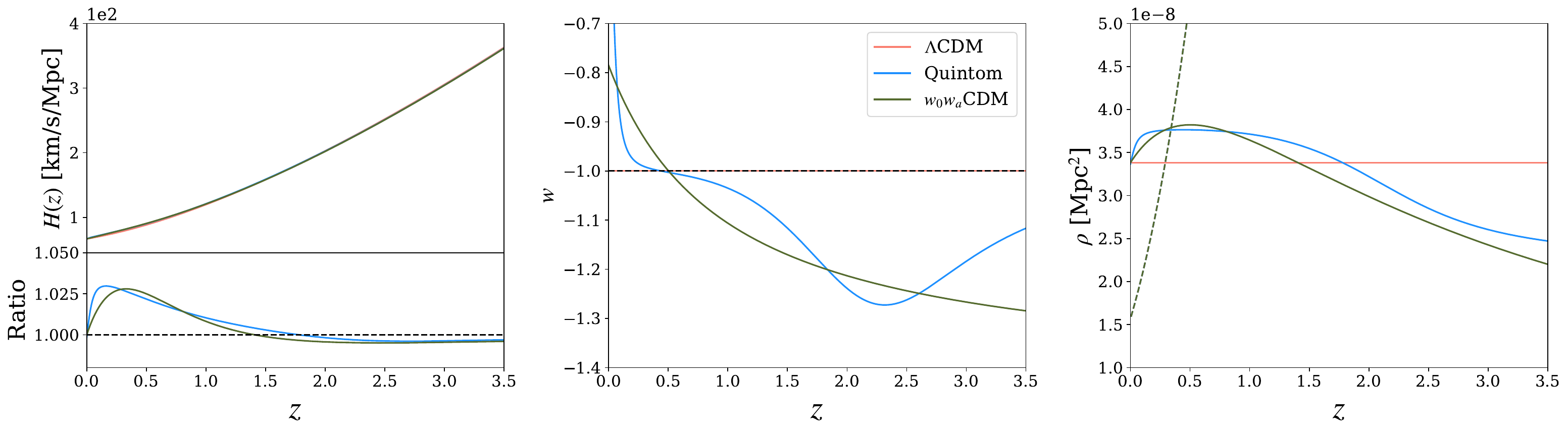} 
    \caption{Left: plot of the Hubble function assuming a $\Lambda$CDM (pink), quintom (blue) or $w_0w_a$CDM (olive) model (top panel), and its ratio with respect to the $\Lambda$CDM model (bottom panel). Middle: plot of the evolution of the equation of state parameter $w$. Right: plot of the dark energy density $\rho_{\rm DE}$ in solid lines, and the matter energy density $\rho_{\rm m}$ is dashed lines. We focus on the redshift range of $0 < z < 3.5$ probed by the data. } \label{fig:background} 
\end{figure*}

We illustrate in Fig.~\ref{fig:background} the background evolution of three cosmological models: $\Lambda$CDM, the quintom model, and a dynamical dark energy model parametrised by $w_0w_a$CDM. We adopt a fiducial $w_0w_a$CDM cosmology with best-fit parameters $w_0=-0.752$ and $w_a=-0.86$ from \citet{2025arXiv250314738D} using the DESI+CMB+DESY5 dataset, together with the best-fit quintom parameters $(V_0,s,\phi_{\rm ini},\psi_{\rm ini})$ obtained from our MCMC analysis (see Table~\ref{tab:bestfit}).

The quintom model exhibits a background evolution qualitatively similar to that of $w_0w_a$CDM. During the phantom dominated phase, the dark energy density grows before subsequently decreasing once quintessence domination begins at $z\sim0.5$. Since $\rho_{\rm DE}<\rho_\Lambda$ up to $z\sim1.5$, the expansion rate initially proceeds more slowly than in $\Lambda$CDM. At intermediate redshifts ($0.5\lesssim z\lesssim1.5$), phantom domination enhances the dark energy density and temporarily increases the expansion rate. As the quintessence component later takes over, $\rho_{\rm DE}$ decreases once more, leading to a suppression in $H(z)$. These variations in the expansion history have important consequences for matter clustering and the growth of large scale structure, which we investigate in the following section. Furthermore, dark energy domination commences earlier in both the quintom and $w_0w_a$CDM cosmologies than in the standard $\Lambda$CDM scenario.

\section{Perturbation Analysis}\label{sec:perturb}

Models that give a varying dark energy $w$ alter the expansion history and consequently, the evolution of the Newtonian gravitational potentials $\Phi$ and $\Psi$ with respect to a cosmological constant framework. This would in turn impact the clustering of dark matter overdensities, manifesting in a change in the LSS observables such as the matter power spectrum and the CMB. In this section we shall conduct a preliminary analysis of the growth of structure in a quintom type-B model, by considering both the perturbations of dark matter and the quintessence and phantom scalar fields. 

We begin with the perturbed metric in the conformal Newtonian gauge,
\begin{equation}\label{eq:perturb_metric}
    ds^2=a(\tau)^2\left[(1+2\Psi)\,d\tau^2-(1-2\Phi)\,dx^idx_i \right],
\end{equation}
where $\tau$ is conformal time, $\Psi$ is the Newtonian gravitational potential, $\Phi$ is the spatial curvature potential, and we have assumed $\Psi=\Phi$ and $\Psi'=\Phi'$ due to the absence of anisotropic stress. Defining the density perturbation of a fluid as $\delta T^0_0=\delta\rho$, its pressure perturbation as $\delta T^i_i=\delta p$ and its velocity divergence $\theta=\frac{ik^j\delta T_j^0}{\rho+p}$, we can write down the $(00)$, $(0i)$ and $(ij)$ components of the perturbed Einstein Field Equations, which relate the evolutions of the potentials to the fluid overdensity and pressure perturbations \citep{1995ApJ...455....7M},
\begin{align}
    \label{eq:efe_perturb00}
    (00)&: k^2\Phi-3\mathcal{H}(\Phi'+\mathcal{H}\Psi)=4\pi Ga^2\delta\rho\,,\\ 
    (0i)&: \Phi'+\mathcal{H}\Psi=4\pi G(\bar{\rho}+\bar{p})\,\theta \frac{a^2}{k^2}\,,\\
    \label{eq:efe_perturb_ij}
    (ij)&:\Phi''+3\mathcal{H}\Phi'+(2\mathcal{H}'-\mathcal{H}^2)\,\Phi=\frac{4}{3}\pi Ga^2\delta p\,. 
\end{align}
Here the prime denotes a derivative with respect to $\tau$, $\mathcal{H}=\frac{a'}{a}$, and the bar denotes the background quantity of the density and pressure. 
From the perturbed part of the energy-momentum conservation equation $\nabla_\mu T^{\mu\nu}=0$, we obtain the evolution of the fluid overdensity $\delta\equiv\delta\rho/\bar{\rho}$ and velocity divergence,
\begin{align}
    \delta' &=-(1+w)\,(\theta-3\Phi')-3\mathcal{H}\left(\frac{\delta p}{\delta\rho}-w\right)\delta\;,\label{eq:dot_delta}\\
    \theta'&=-\mathcal{H}\,(1-3w)\,\theta-\frac{w'}{1+w}\theta+\frac{\delta p}{\delta\rho}\frac{1}{1+w}k^2\delta+k^2\Psi\label{eq:dot_theta}\,.
\end{align}

Depending on the fluid in question (baryons, cold dark matter, quintessence or phantom field), the gravitational potentials $\Phi$ and $\Psi$, which source $\delta p$ and $\delta\rho$, will evolve and directly impact the dynamics of clustering and large scale structure formation. 

\subsection{Matter Perturbations}
Since matter (baryons and cold dark matter) is pressureless, we have $\delta p=0, w=0$ and thus Eqs. \eqref{eq:efe_perturb00} and \eqref{eq:dot_delta} reduce to the linearised scale-independent growth equation at subhorizon scales ($k\ll a\mathcal{H}$), unchanged from $\Lambda$CDM,
\begin{equation}\label{eq:dm_poisson}
    \delta''_{\rm m}+2\mathcal{H}\delta'_{\rm m}-4\pi G a^2\rho_{\rm m}\delta_{\rm m} = 0\,.
\end{equation}

\begin{figure}
    \centering
    \includegraphics[width=\linewidth]{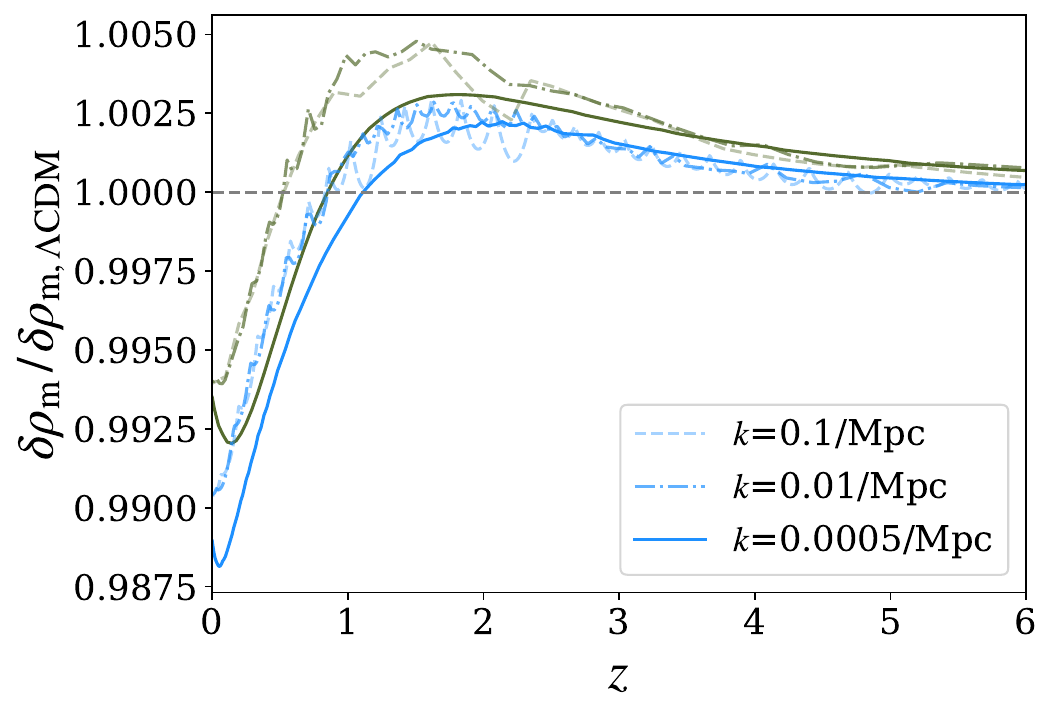}
    \caption{Ratio of the evolution of matter perturbations $\delta\rho_{\rm m}$ for a $w_0w_a$CDM model (olive) and a quintom model (blue) with respect to $\Lambda$CDM at different $k$ scales, $k=\{5\times10^{-4},1\times10^{-2},1\times10^{-1}\}$ /Mpc. Note that the oscillations at low $k$ arise from numerical errors due to redshift interpolation.}
    \label{fig:dm_perturb}
\end{figure}

We modify the Einstein-Boltzmann code \textsc{CLASS} \citep{2011JCAP...07..034B} to include a two-field quintom model with a hyperbolic tangent potential as defined in Eq. \eqref{eq:v_tanh}.  In Fig. \ref{fig:dm_perturb} we plot the matter perturbations at different $k$ scales, for the same cosmological parameter values employed in the previous section to derive the background quantities. We see that the general trend of modifications to $\delta_{\rm m}$ with respect to $\Lambda$CDM remain largely scale-independent. At early times, a reduced rate of expansion (as demonstrated in Fig. \ref{fig:background}) gives rise to enhanced clustering, before phantom domination eventually leads to $\rho_{\rm DE}>\rho_{\Lambda}$ and thus increased expansion, subsequently suppressing the growth of matter overdensities at late times. The large scale perturbations, having entered the horizon later, have less time to grow during the phantom dominated phase before being suppressed, hence resulting in the smallest amplitude of $\delta_{\rm m}$ at the largest scales (for e.g. at $k=5\times10^{-4}$/Mpc). We also see we see an upturn of $\delta\rho_{\rm m}$ at very late time, as they respond to the quintessence dominated phase where expansion slows and clustering once again is enhanced.  

\subsection{Scalar Field Perturbations}
From Eqs. \eqref{eq:p_rho_phi} and \eqref{eq:p_rho_psi}, we can derive the density and pressure perturbations for a quintessence scalar field,  
\begin{align} 
\delta\rho_\phi&=\frac{1}{a^2}\left(\bar{\phi}'\delta\phi'-\bar{\phi}'^{2}\,\Psi\right)+V_{\phi}\,\delta\phi\,,\\ 
\delta p_\phi&=\frac{1}{a^2}\left(\bar{\phi}'\delta\phi'-\bar{\phi}'^{2}\,\Psi\right)-V_{\phi}\,\delta\phi\,. 
\end{align} 
The perturbed Klein-Gordon equation is then given by 
\begin{equation} \label{eq:perturb_kg_phi}
\delta\phi''+2\mathcal{H}\delta\phi'+(k^2+a^2V_{\phi\phi})\,\delta\phi-3\Phi'\phi'=0\,. \end{equation}
The metric perturbations directly source the evolution of the scalar field perturbations through the perturbed Klein-Gordon equation. Conversely, the scalar field perturbations contribute to the perturbed energy-momentum tensor through $\delta\rho_\phi$ and $\delta p_\phi$, thereby sourcing the evolution of the gravitational potentials via the perturbed Einstein equations in Eqs. \eqref{eq:efe_perturb00}$-$\eqref{eq:efe_perturb_ij}. The dynamics of the scalar field and metric perturbations are therefore intrinsically coupled.

For a phantom field $\psi$, the relevant equations are
\begin{align}
    \delta\rho_\psi&=-\frac{1}{a^2}\left(\bar{\psi}'\delta\psi'-\bar{\psi}'^{2}\,\Psi\right)+V_{\psi}\,\delta\psi\,,\label{eq:delta_rho_psi}\\
    \delta p_\psi&=-\frac{1}{a^2}\left(\bar{\psi}'\delta\psi'-\bar{\psi}'^{2}\,\Psi\right)-V_{\psi}\,\delta\psi\,,
\end{align}
with the perturbed KG equation given by
\begin{equation}\label{eq:perturb_kg_psi}
\delta\psi''+2\mathcal{H}\delta\psi'+(k^2-a^2V_{\psi\psi})\,\delta\psi-3\Phi'\psi'=0\,.
\end{equation}
Since we have assumed minimal coupling between the fields, they obey these evolution equations separately without inducing mixing of terms. The effective dark energy overdensity and velocity divergence of a two-field quintom system is then the sum of that of each field $i\in[\phi,\psi]$ \citep{2005PhRvD..72l3515Z}
\begin{align}
    \delta_{\rm DE}&=\frac{\Sigma_i\,\rho_i\,\delta_i}{\Sigma\,\rho_i}\,,\\
    \theta_{\rm DE}&=\frac{\Sigma_i\,(\rho_i+p_i)\,\theta_i}{\Sigma_i\,(\rho_i+p_i)}\,.
\end{align}

\begin{figure*}
    \centering
    \includegraphics[width=\linewidth]{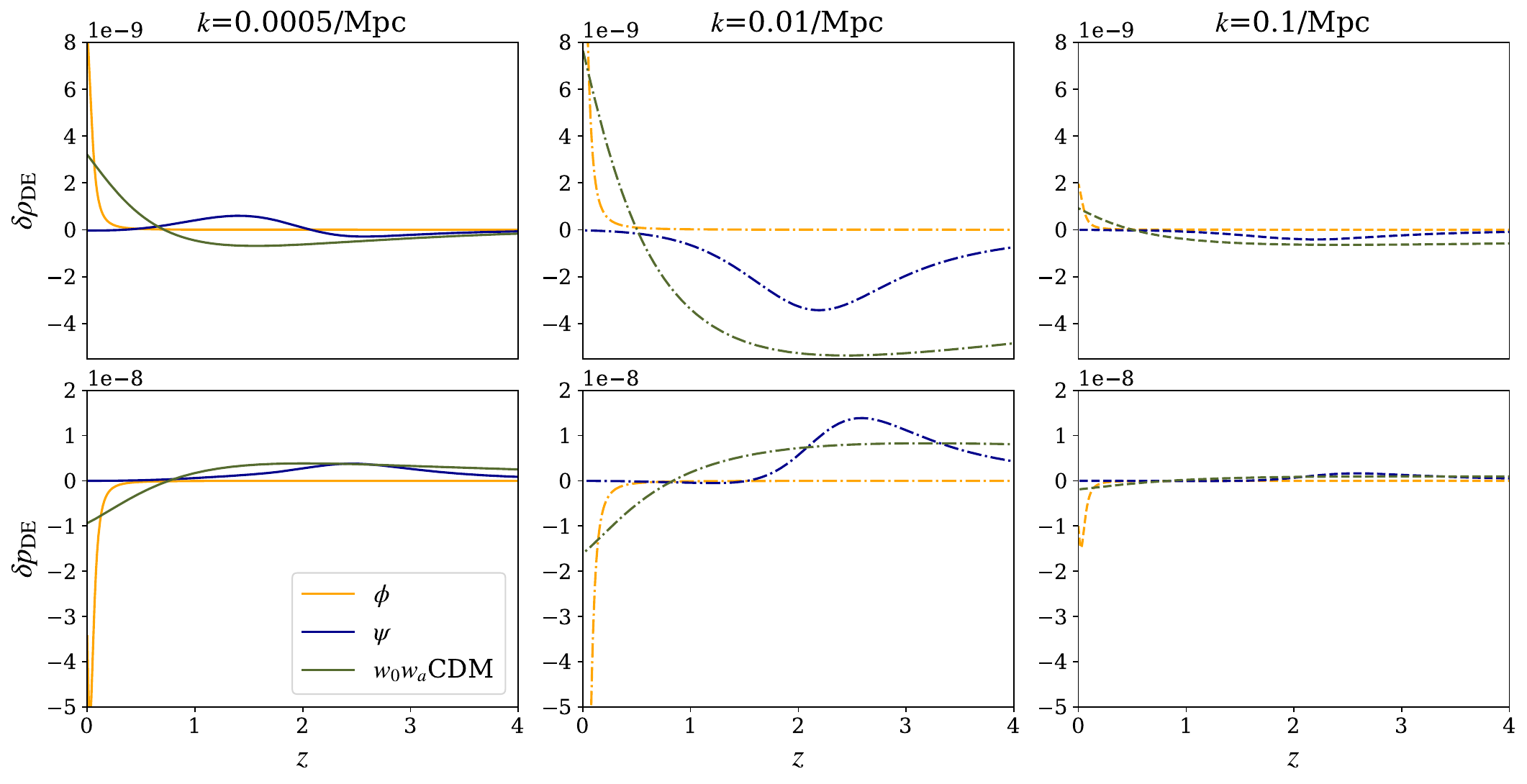}
    \caption{Top: The dark energy fluid density perturbations $\delta\rho_{\rm DE}$ in a $w_0w_a$CDM (olive) and a quintom model (yellow: quintessence field contribution, blue: phantom field contribution), for $k$ scales of $k=0.0005$/Mpc (left, solid lines), $k=0.01$/Mpc (middle, dashed dotted lines) and $k=0.1$/Mpc (right, dashed lines). Bottom: similar plot for the pressure perturbations $\delta p_{\rm DE}$.}
    \label{fig:dscf_perturb}
\end{figure*}

 \begin{figure*}
    \centering
    \includegraphics[width=\linewidth]{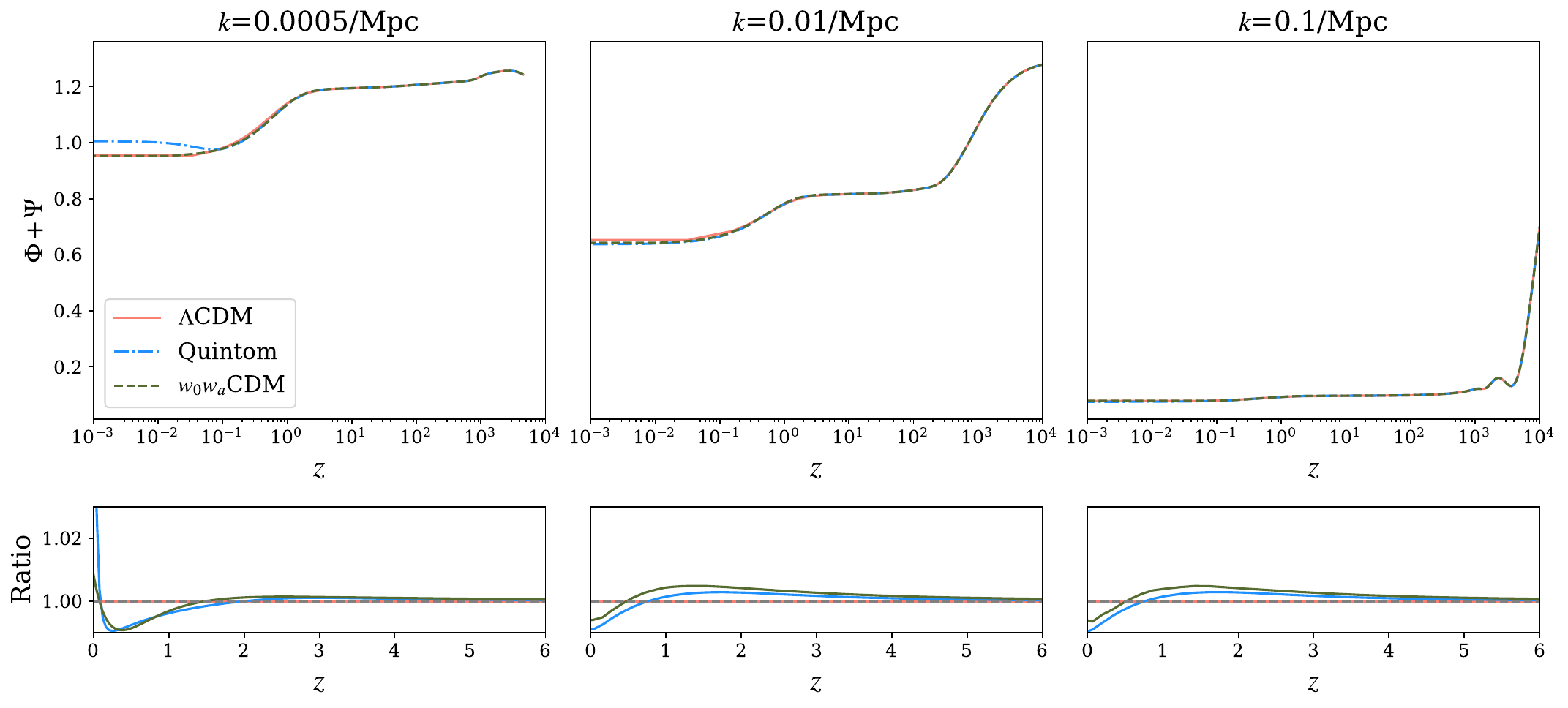}
    \caption{Plot of the evolution of the gravitation potentials $\Phi+\Psi$ at increasing scales from left to right. The top panels show their absolute magnitude for the three cosmological models: $\Lambda$CDM (pink), quintom (blue) and $w_0w_a$CDM (olive). In the bottom panels we plot their ratios with respect to the $\Lambda$CDM model, particularly focusing on the low redshift range of $0<z<6$. Note the different scales used for the redshift axis between the top and bottom plots.}
    \label{fig:phi_psi}
\end{figure*}

In Fig.~\ref{fig:dscf_perturb}, we show the evolution of the dark energy density and pressure perturbations, $\delta\rho_{\rm DE}$ and $\delta p_{\rm DE}$, for both the quintom and $w_0w_a$CDM models at several representative scales. We see that the quintom model reproduces the broad behaviour of the effective dark energy perturbations obtained in the fluid $w_0w_a$CDM parametrisation. In the scalar field description, dark energy perturbations arise from the field fluctuations $\delta\phi$ and $\delta\psi$, and their relative contributions evolve as the background dynamics transition from phantom to quintessence domination.

Furthermore, two main trends emerge, specifically with respect to redshift and scale. At early times, when the phantom field dominates the dark energy sector, the perturbations are primarily sourced by $\delta\psi$ (blue curves). As the phantom field approaches the plateau of the potential and slows, the amplitude of $\delta\rho_\psi$ decreases, while the quintessence perturbation $\delta\rho_\phi$ begins to grow as the field rolls down the potential and becomes dynamically important. The transition between these regimes occurs at approximately the same redshift as the sign change in the effective dark energy perturbations of the $w_0w_a$CDM model, indicating that underlying scalar field dynamics of a quintom framework is able to motivate the overall behaviour of a dark energy fluid.

The sign of $\delta\rho_\psi$ is sensitive to the competing contributions between the kinetic and potential terms in Eq.~\eqref{eq:delta_rho_psi}. For the hyperbolic tangent potential considered here, $V_{,\psi}<0$ and $\delta\psi>0$, implying that the potential contribution $V_{,\psi}\delta\psi$ is always negative. Whether $\delta\rho_\psi$ is positive or negative therefore depends on the kinetic term involving $\delta\psi'$. This behaviour is most clearly seen on the largest scales ($k=0.0005\,{\rm Mpc}^{-1}$), where $\delta\rho_\psi$ becomes positive at intermediate redshifts before decaying towards zero at late times. On the other hand, the perturbations of the quintessence field is always positive, since $\delta\phi'>0$ at all scales and redshifts.

Secondly, we see a pronounced scale-dependence of $\delta\rho_{\rm DE}$ and $\delta\rho_{\rm DE}$ in both the $w_0w_a$CDM model and the $\phi$ and $\psi$ fields of the quintom model, whereby their amplitude is strongest at intermediate scales. This is once again a direct consequence of the interplay between the gradient term $\frac{k^2}{a^2}$ and potential term $V_{\phi\phi}$ appearing in the perturbed KG equation of Eqs. \eqref{eq:perturb_kg_phi} and \eqref{eq:perturb_kg_psi}. 
 
 Analogous to the evolution of $\delta\rho_{\rm m}$, the evolution of scalar field perturbations depends strongly on horizon crossing. Large scale modes remain super-horizon for longer periods before re-entering the horizon as the Universe expands, and therefore experience less sustained growth. Conversely, small scale modes re-enter the horizon earlier, but their evolution is efficiently suppressed by Hubble friction and gradient pressure. As a result, the perturbation amplitude is maximised at intermediate scales, such as $k\sim0.01$/Mpc. These perturbations grow most significantly when the corresponding field component (phantom or quintessence) becomes dynamically dominant. The larger magnitude of the perturbations can be largely attributed to a larger $w_{\rm DE}$ in the quintom model during quintessence domination (middle panel of Fig. \ref{fig:background}). Nevertheless, the contribution of $\delta\rho_{\rm DE}$ to the total fluid perturbation is subdominant to that of $\delta\rho_{\rm m}$, which we have verified to be three orders of magnitude larger. Consequently, observable modifications to large scale structure are expected to arise primarily from changes in the matter perturbation growth induced by the altered expansion history, rather than from perturbations of the scalar fields themselves.

The density and pressure perturbations of the matter and dark energy subsequently source the evolution of the gravitational potentials $\Phi$ and $\Psi$, for which their sum is plotted in Fig. \ref{fig:phi_psi}. Once again, we observe a clear scale dependence: larger scale modes remain frozen during the matter-dominated era for longer durations, appearing as extended plateaus in the evolution of $\Phi+\Psi$, before evolving rapidly once dark energy domination begins. Smaller-scale modes, such as $k=0.1$/Mpc, enter the horizon earlier and undergo more efficient damping, leading to a reduced late-time amplitude in the potentials, as seen in the rightmost panel.

As expected, modifications to the potentials are predominantly driven by the matter overdensity $\delta\rho_{\rm m}$, and therefore closely tracks the behaviour shown in Fig.~\ref{fig:dm_perturb}. At early times, enhanced matter clustering deepens the gravitational potential wells, increasing the amplitude of $\Phi+\Psi$. At later times, once $\rho_{\rm DE}>\rho_\Lambda$, the accelerated expansion suppresses structure growth and leads to a decay of the potentials. The late-time enhancement in large scale matter perturbations is correspondingly reflected in the evolution of $\Phi+\Psi$.

\subsection{Impact on Observations}

\begin{figure}
    \centering
    \includegraphics[width=\linewidth]{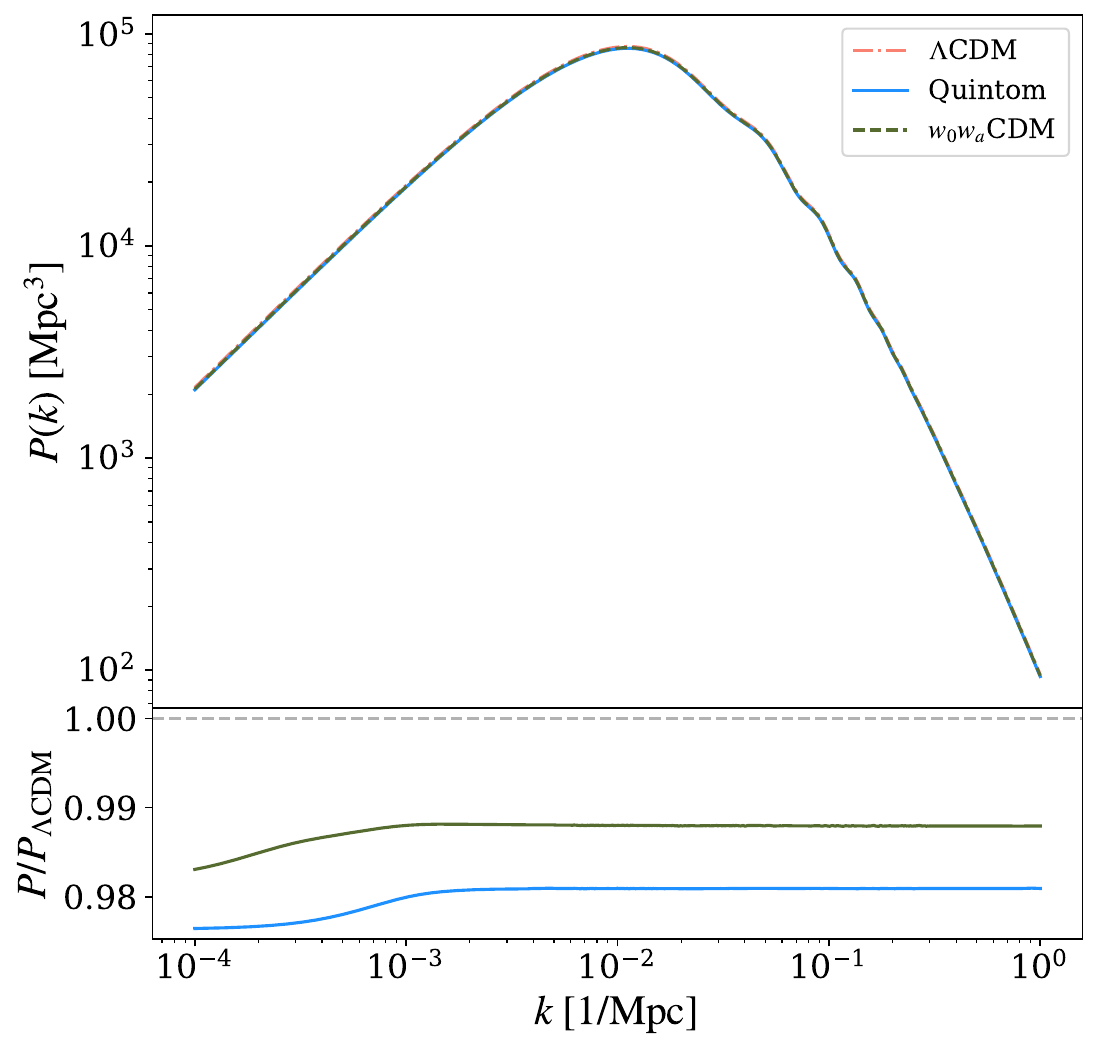}
    \caption{Upper panel: Plot of the linear matter power spectrum $P(k)$ for a $\Lambda$CDM (pink), quintom (blue) and $w_0w_a$CDM (olive) model. Lower panel: Plot of their ratios with respect to $\Lambda$CDM.}
    \label{fig:pk}
\end{figure}

Finally, we examine the observational implications of a phantom-to-quintessence quintom cosmology for LSS formation and the CMB. In Fig.~\ref{fig:pk}, we plot the linear matter power spectrum for the three cosmological models and show their ratios relative to $\Lambda$CDM. We find that matter clustering is suppressed across all scales in both the quintom and $w_0w_a$CDM scenarios, consistent with the behaviour of $\delta\rho_{\rm m}$ shown in Fig.~\ref{fig:dm_perturb}. The suppression is strongest on large scales, reflecting the enhanced sensitivity of near-horizon modes to modifications in the expansion history. However the overall effect remains marginal, on the order of 1--2\%. Nonetheless it should be noted that we have only considered the linear power spectrum thus far, and leave the development of the nonlinear prescription for future work. With an accurate modelling of the nonlinear power spectrum, we anticipate better discrimination between cosmological constant, dynamical dark energy and quintom models. 

The most significant imprint on the CMB arises through the late-time Integrated Sachs--Wolfe (ISW) effect, as previously noted by \citet{2003PhRvL..91g1301C,2005PhRvD..72l3515Z}. During the epoch of dark energy domination, accelerated expansion causes the gravitational potentials to evolve with time. CMB photons traversing these decaying potential wells consequently gain or lose energy, generating additional temperature anisotropies. The ISW contribution is proportional to the line-of-sight integral of $\Phi'+\Psi'$, and therefore primarily affects the low-$\ell$ multipoles corresponding to large angular scales.

\begin{figure}
    \centering
    \includegraphics[width=\linewidth]{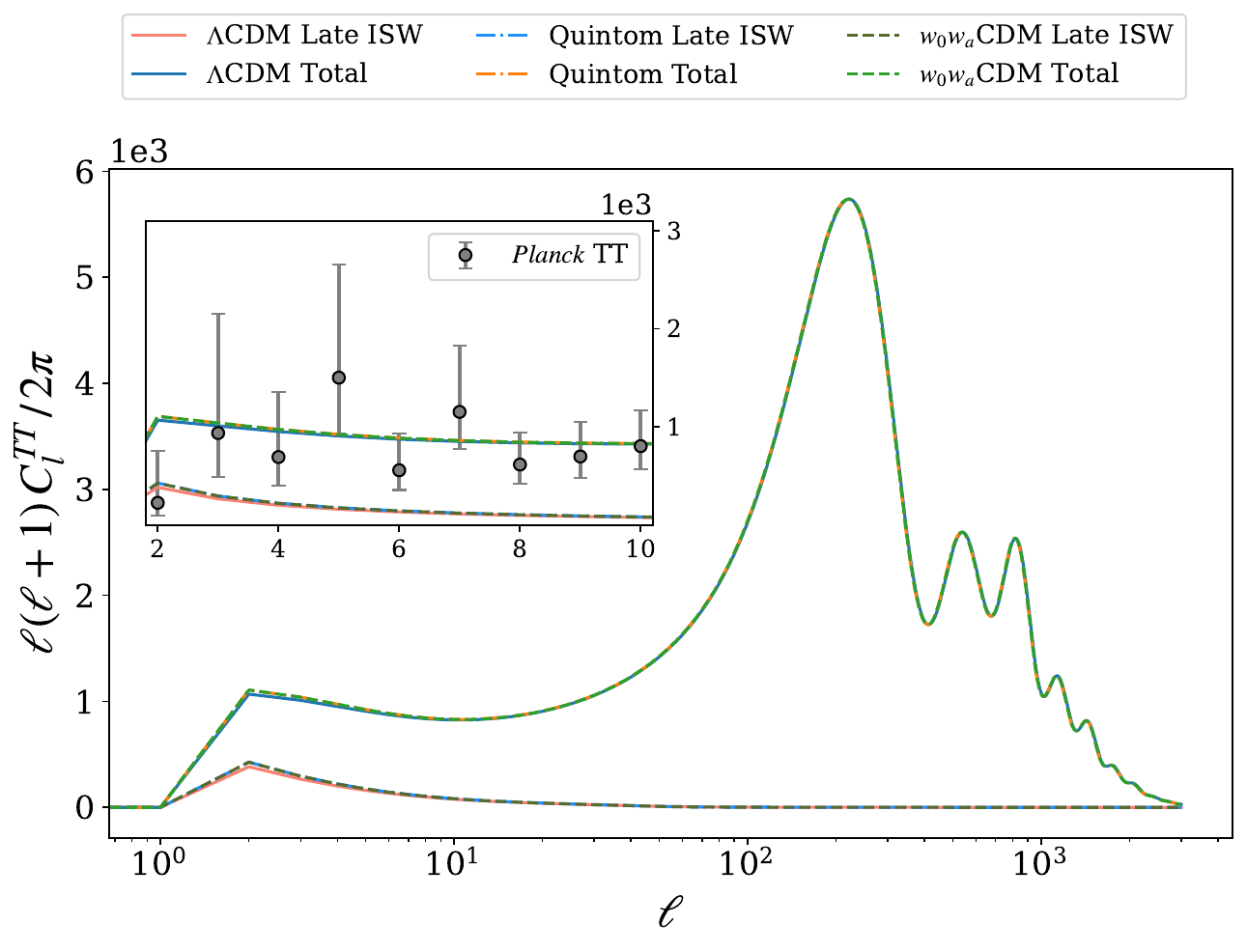}
    \caption{Plot of the CMB power spectrum $C_\ell^{TT}$, where we plot the total lensed spectrum and isolate the late-time ISW contribution for the three models. The inset plot shows a zoom-in at the large scales of $2\leq \ell\leq 10$, where we include the $C_\ell^{TT}$ data points from \textit{Planck} 2018.}
    \label{fig:cl}
\end{figure}

As shown in Fig.~\ref{fig:cl}, cosmologies with an evolving dark energy equation of state exhibit a mild enhancement in the large-scale CMB anisotropies relative to $\Lambda$CDM, owing to the increased ISW contribution. The phantom-to-quintessence transition induces a more pronounced evolution of the gravitational potentials $\Phi+\Psi$, leading to a correspondingly stronger ISW signal. However, similar to the matter power spectrum, the resulting deviations remain at the percent level. Such signatures would therefore be difficult to distinguish observationally given the current precision limits of CMB and LSS measurements, particularly in the presence of cosmic variance on large angular scales. This is highlighted in the inset plot of Fog. \ref{fig:cl}, where the error bars of the temperature anisotropies from the  \textit{Planck} dataset are much larger than deviations in their modelling due to the different models.  

\section{Inference} \label{sec:mcmc}

Using our modified Boltzmann code, we perform a full Bayesian inference analysis to determine whether a phantom-to-quintessence transition motivated by a quintom model is supported by current cosmological data. The analysis was carried out using \texttt{Cosmosis} \citep{2015A&C....12...45Z}, with the \texttt{Polychord} \citep{2015MNRAS.450L..61H,2015MNRAS.453.4384H} nested sampler, which enables efficient sampling in high-dimensional parameter spaces. We vary the standard cosmological parameters $\omega_{\rm b},\;\omega_{\rm c},\;n_{\rm s},\;\tau,\;\ln(10^{10}A_{\rm s}),$ and, for the quintom cosmology, additionally sample the scalar field parameters $V_0,\;s,\;\phi_{\rm ini},\;\psi_{\rm ini}$. Unlike phenomenological dark energy parametrisations, the background evolution in the quintom framework is determined through its scalar field dynamics. Specifically, the energy densities $\rho_\phi$ and $\rho_\psi$ are evolved directly from the Klein-Gordon equations, Eqs. \eqref{eq:kg_phi} and \eqref{eq:kg_psi}, while the Hubble parameter is obtained from the Friedmann equation, Eq.~\eqref{eq:friedmann}. Consequently, $H_0$ is treated as a derived parameter rather than an independently sampled one in a quintom model. Throughout this work, we assume a spatially flat Universe and fix the effective number of relativistic neutrino species to $N_{\rm eff}=3.044$ \citep{2019JCAP...07..014G}.
We adopt the same observational datasets used by \citet{2025arXiv250314738D}, combining constraints from both early- and late-time cosmological probes at the background and perturbation levels. These include baryon acoustic oscillation (BAO) measurements from the DESI survey, comprising the Bright Galaxy Sample (BGS), Luminous Red Galaxies (LRGs), Emission Line Galaxies (ELGs), quasars (QSOs), and Lyman-$\alpha$ forest samples; the \textit{Planck} 2018 CMB TT, TE, and EE likelihoods \citep{2020A&A...641A...6P}; and the Atacama Cosmology Telescope \citep[ACT;][]{2024ApJ...962..113M,ACT:2023dou} DR6 lensing likelihood \citep{2022JCAP...09..039C}. These datasets are further combined with Type Ia supernova measurements from both the Dark Energy Survey Year 5 release \citep[DESY5;][]{2024ApJ...973L..14D} and the Pantheon+ survey \citep{2022ApJ...938..113S}.

\section{Results}\label{sec:results}

\begin{figure*}
    \centering
    \includegraphics[width=\linewidth]{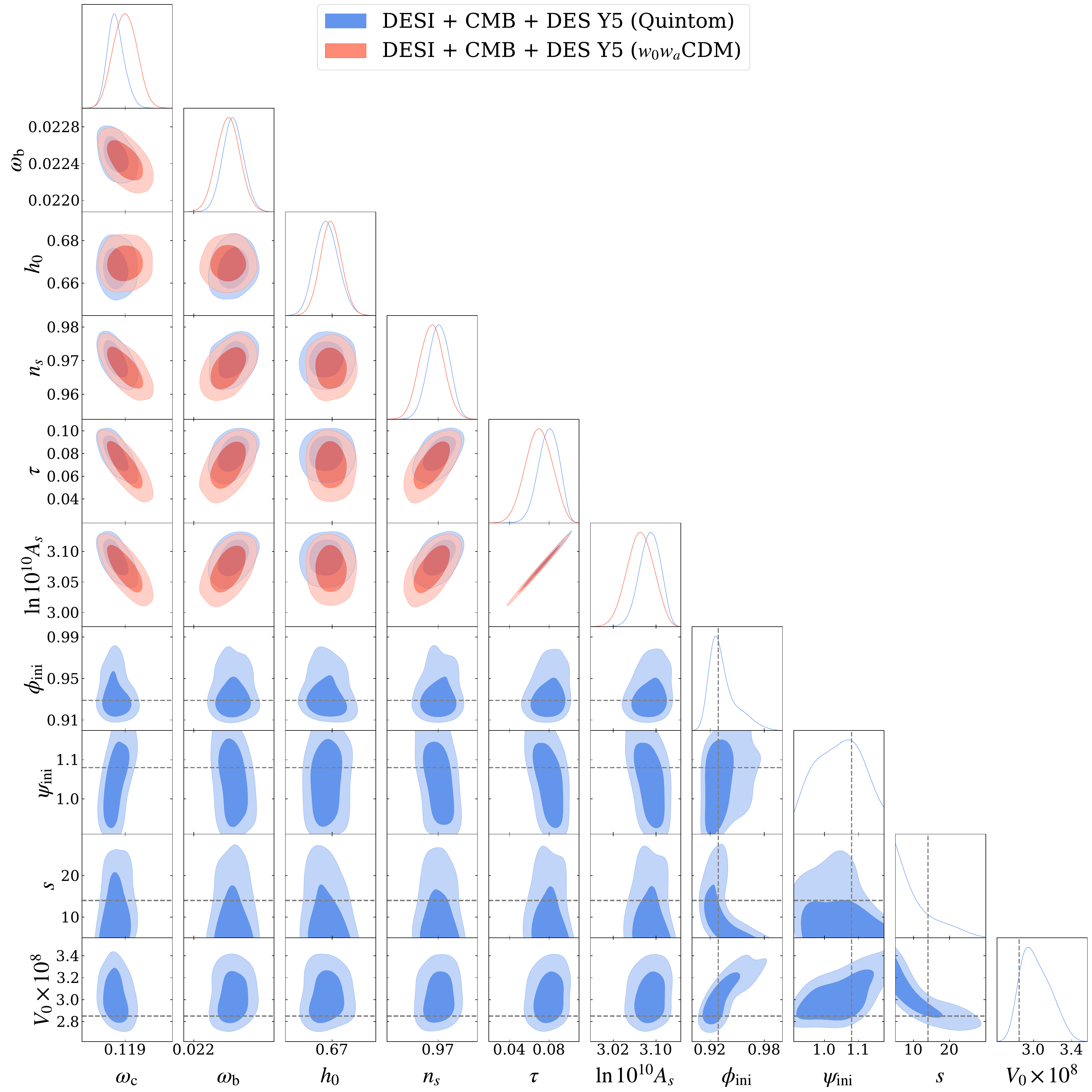}
    \caption{2D marginalised posterior distributions of the cosmological parameters for the $w_0w_a$CDM parameterisation (red) and the quintom model (blue), for a DESI+CMB+DESY5 dataset. We denote the MAP values of the quintom model parameters in grey dashed lines.}
    \label{fig:contour_des}
\end{figure*}

\begin{figure*}
    \centering
    \includegraphics[width=\linewidth]{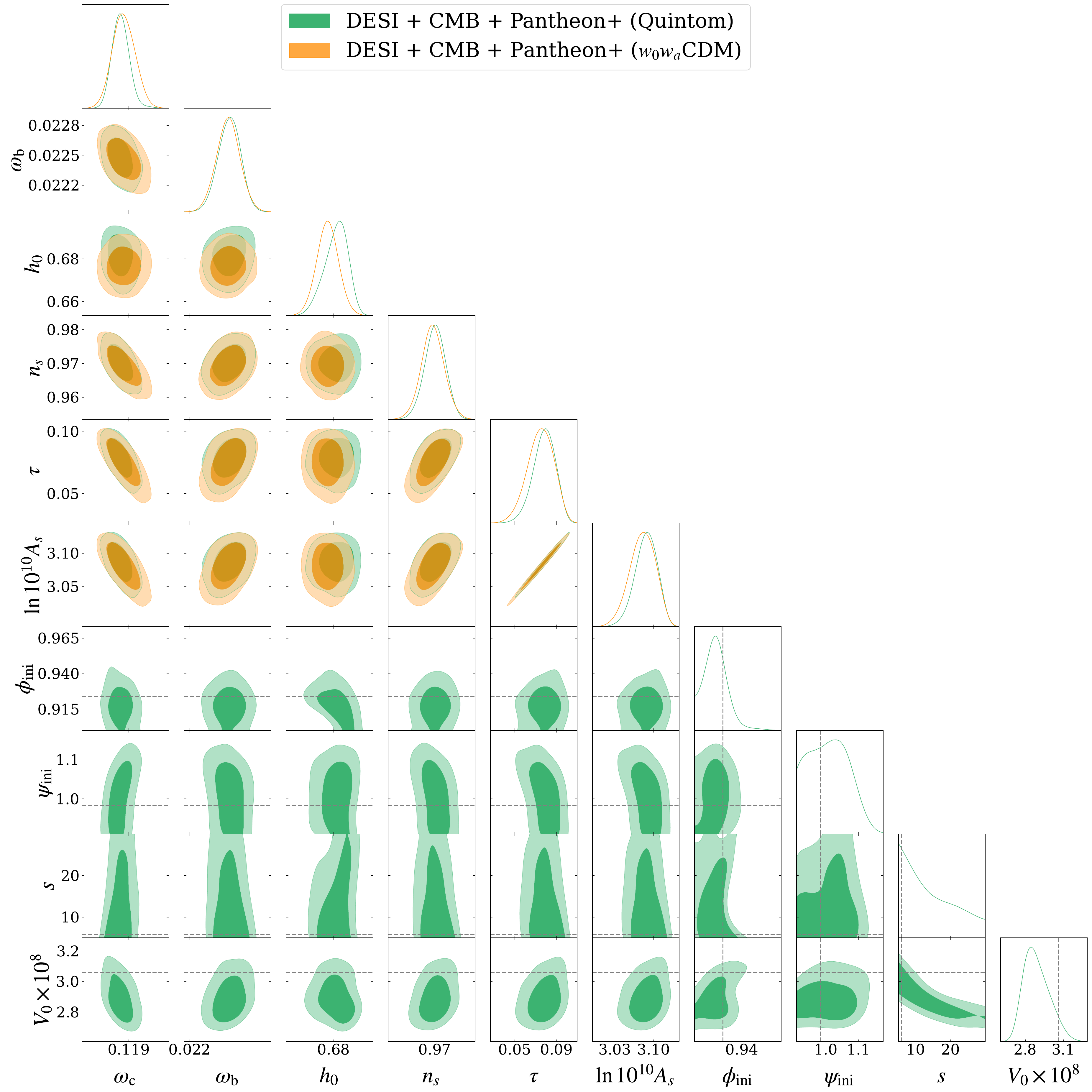}
    \caption{2D marginalised posterior distributions of the cosmological parameters for the $w_0w_a$CDM parameterisation (orange) and the quintom model (green), for a DESI+CMB+Pantheon+ dataset. We denote the MAP values of the quintom model parameters in grey dashed lines.}
    \label{fig:contour_pantheon}
\end{figure*}

In Fig. \ref{fig:contour_des} and Fig. \ref{fig:contour_pantheon} we present the full 2D marginalised posterior distributions of the cosmological parameters for the DESI+CMB+DESY5 and DESI+CMB+Pantheon+ datasets respectively, for both a quintom and $w_0w_a$CDM model. As established in the previous section, scalar perturbations have a marginal impact on the CMB power spectra, so our constraints are expected to be driven primarily by background quantities — namely the expansion history encoded in cosmological distances measured from BAO and SNe1a. The CMB instead anchors the early-Universe parameters $n_{\rm s}$, $\tau$, $\ln(10^{10}A_{\rm s})$, and $\omega_{\rm b}$, which should not vary appreciably between models. This behaviour is confirmed by the 1D posteriors, which are largely consistent across both models, with only a marginally lower value of $\omega_{\rm c}$ preferred in the quintom case.

Furthermore, we note that the posterior distributions of the quintom model parameters are highly non-Gaussian and exhibit strong degeneracies, particularly among $V_0$, $\phi_{\rm ini}$, and $s$. We therefore choose to employ the Maximum a Posteriori (MAP) estimator when quoting the best-fit parameter values, rather than the conventional weighted mean obtained through marginalisation over the parameter space. In a high-dimensional and non-Gaussian parameter space, such marginalisation can introduce prior volume effects, whereby the resulting 1D marginalised distribution becomes biased and unrepresentative of the underlying multivariate posterior \citep{2022PhRvD.106f3506G}. This is a well-known issue in cosmological parameter inference \citep{2021A&A...646A.129J,2021arXiv210513548K,2023JCAP...01..028C}, and has motivated the use of alternative summary statistics in Bayesian analyses, such as the MAP estimator. 

The MAP is defined as the single point in the multivariate parameter space that maximises the posterior probability,
\begin{equation} 
\mathbf{\Theta}_{\rm MAP}=\text{argmax}\;p(\Theta|d)\,. 
\end{equation}
Since these parameter values are taken directly from the sampled chain without any marginalisation, this approach mitigates projection effects and provides a more faithful indication of the preferred region of the full posterior distribution. We note, however, that the MAP estimator can be unstable due to sensitivity to sampling noise, particularly for sparsely sampled chains \citep{des-muir}.

To quantify parameter uncertainties, we estimate the 1D highest posterior density (HPD) interval and quote the 68\% HPD region as the corresponding confidence interval. This is obtained by first fitting a smooth probability density function for each parameter using Gaussian kernel density estimation (KDE), which helps mitigate sampling noise arising from a limited number of samples. The density values are then ranked in descending order, and the highest density region containing 68\% of the total posterior probability mass is identified as the HPD interval.

In Table~\ref{tab:bestfit}, we report the best-fit values and their associated 68\% confidence intervals for each setup considered. We can see that the best-fit values in the quintom model do not necessarily coincide with the peaks of the 1D marginal posterior distributions shown in Figs.~\ref{fig:contour_des} and \ref{fig:contour_pantheon}, directly reflecting the non-Gaussian and highly degenerate structure of the parameter space. This discrepancy arises because the MAP identifies the single point within the joint posterior distribution with the highest posterior probability, whereas the peak of a 1D marginal distribution is a projection obtained 
by integrating over all other parameters. Furthermore, the 68\% HPD regions are generally much broader than the confidence intervals inferred from the marginalised posterior distributions, a behaviour that is expected given the finite chain length and the degree of smoothing applied in the KDE step.

\begin{table}
    \centering
    \fontsize{6.3}{11}\selectfont
\begin{tabular}{|c|c|c|c|c|}
\hline
    \textbf{Parameter} & 
         \makecell{\textbf{DESI}\\\textbf{+CMB}\\ \textbf{+DESY5}\\ \textbf{(Quintom)}} & 
         \makecell{\textbf{DESI}\\\textbf{+CMB} \\ \textbf{+Pantheon+} \\\textbf{(Quintom)}} & \makecell{\textbf{DESI}\\\textbf{+CMB}\\\textbf{+DESY5}\\\textbf{(}$w_0w_a$\textbf{CDM)}} & 
         \makecell{\textbf{DESI}\\\textbf{+CMB}\\ \textbf{+Pantheon+} \\\textbf{(}$w_0w_a$\textbf{CDM)}}\\
         \hline
         $\omega_{\rm c}$& 0.119$^{+0.009}_{-0.008}$&0.118$^{+0.011}_{-0.008}$&0.119$\pm0.0011$&0.119$\pm0.0010$\\
         $\omega_{\rm b}$& 0.0224$\pm0.0012$&0.0224$^{+0.0014}_{-0.0015}$&0.0224$\pm0.00014$&0.0225$^{+0.00013}_{-0.00014}$\\
         $n_{\rm s}$&0.969$\pm0.030$&0.972$^{+0.030}_{-0.038}$&0.968$^{+0.0040}_{-0.0041}$&0.969$^{+0.0039}_{-0.0038}$\\
         $\ln{(10^{10}A_{\rm s})}$& 3.079$^{+0.053}_{-0.068}$ &3.097$^{+0.036}_{-0.094}$&3.071$^{+0.0257}_{-0.0249}$&3.080$^{+0.0253}_{-0.0215}$\\
         $\tau$& 0.0746$^{+0.023}_{-0.021}$&0.0842$^{+0.013}_{-0.034}$&0.0696$\pm0.0140$&0.0748$^{+0.0136}_{-0.0116}$\\
         $H_0$& 66.34$^{+3.65}_{-2.96}$&67.45$^{+3.20}_{-2.90}$&66.93$\pm0.55$&67.66$\pm0.59$\\
         $w_0$& $-$ &$-$&$-0.785^{+0.0551}_{-0.0596}$&$-0.870^{+0.0554}_{-0.0551}$\\
         $w_a$&$-$ &$-$&$-0.642^{+0.246}_{-0.212}$&$-0.385^{+0.224}_{-0.197}$\\
         $V_0\times 10^{8}\,[m_{\rm P}^4]$& 2.85$^{+5.17}_{-2.61}$&3.06$^{+2.62}_{-5.66}$&$-$&$-$\\
         $s\,[m_{\rm P}^{-1}]$&14.0$^{+4.18}_{-8.99}$ &5.79$^{+14.15}_{-0.786}$&$-$&$-$\\
         $\phi_{\rm ini}\,[m_{\rm P}]$& 0.929$^{+0.012}_{-0.026}$&0.924$^{+0.012}_{-0.020}$&$-$&$-$\\
         $\psi_{\rm ini}\,[m_{\rm P}]$&1.080$^{+0.109}_{-0.041}$ &0.983$^{+0.121}_{-0.027}$&$-$&$-$\\
         \hline
    \end{tabular}
    \caption{Table of best-fit values and their 1$\sigma$ uncertainties for the cosmological parameters, for both datasets and models studied. Note that we quote the MAP and the 68\% HPD for the quintom models, while we keep the use of the weighted mean and take the 68\% CI of the marginalised posterior distribution in the $w_0w_a$CDM models. Note that our results for the $w_0w_a$CDM models are not necessarily identical to those of the fiducial analysis in \citet{2025arXiv250314738D} due to a difference in analysis setup.}
    \label{tab:bestfit}
\end{table}

\subsection{Comparison to data}
\begin{figure*}    
\centering    
\includegraphics[width=\linewidth]{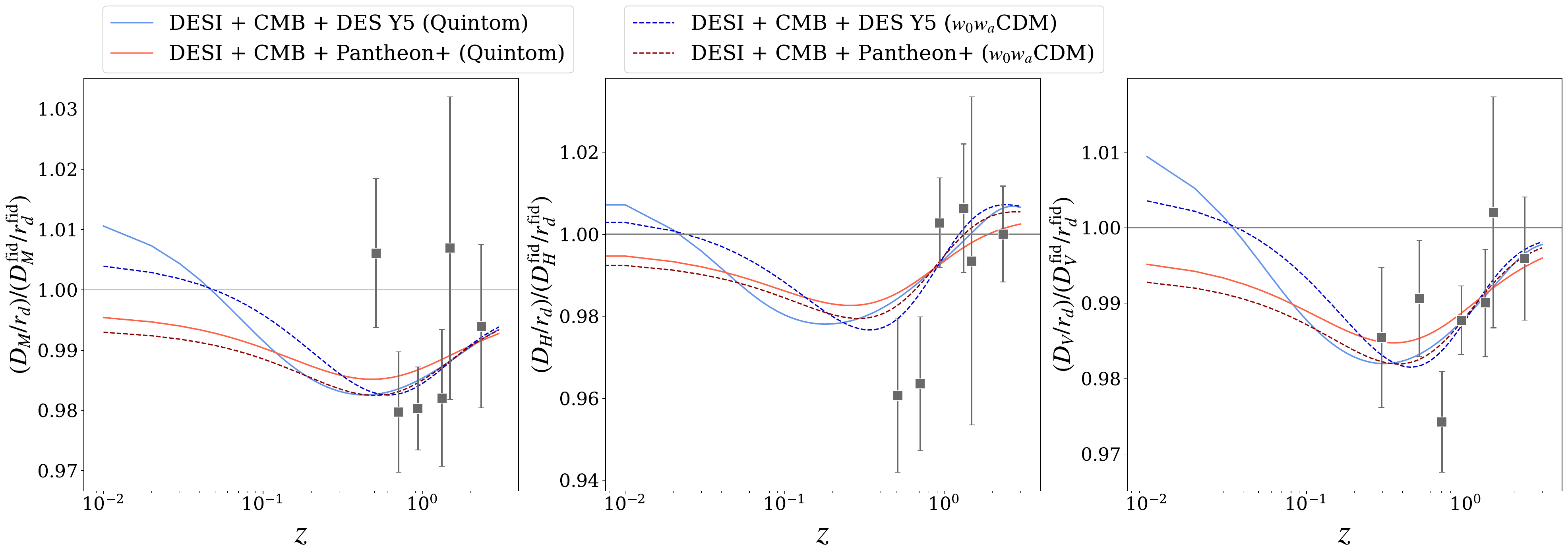}
\caption{Ratio of the transverse comoving BAO distance for the best-fit model of each cosmology (quintom or $w_0w_a$CDM) and dataset combination (DESY5 or Pantheon+) relative to the fiducial value, $(D_M/r_d)/(D^{\rm fid}_M/r^{\rm fid}_d)$ (left panel), where the fiducial cosmology is taken to be the \textit{Planck} 2018 $\Lambda$CDM best-fit model \citep{2020A&A...641A...6P}. Filled grey points show the DESI BAO measurements (corresponding to the first seven cosmological measurements listed in Table IV of \citet{2025arXiv250314738D}) with their $1\sigma$ uncertainties. The middle and right panels show the corresponding ratios for the line-of-sight comoving distance, $(D_H/r_d)/(D^{\rm fid}_H/r^{\rm fid}_d)$, and the isotropic BAO distance, $(D_V/r_d)/(D^{\rm fid}_V/r^{\rm fid}_d)$, respectively.}    
\label{fig:bestfit_distance}
\end{figure*}

Using the best-fit parameter values obtained for each cosmological model and dataset combination, we compute the corresponding cosmological observables for both the quintom and $w_0w_a$CDM scenarios and compare them directly with the observational measurements. In Fig.~\ref{fig:bestfit_distance}, we plot the ratios of the BAO distance measures to the sound horizon scale, namely $D_{\rm M}/r_{\rm d}$, $D_{\rm H}/r_{\rm d}$, and $D_{\rm V}/r_{\rm d}$. We find that the best-fit quintom cosmology successfully reproduces the characteristic behaviour of a $w_0w_a$CDM parametrisation, particularly the decreasing trend relative to the fiducial $\Lambda$CDM prediction exhibited by the BAO data over the redshift range $0.4 \lesssim z \lesssim 3$.

\begin{figure}
    \centering
    \begin{subfigure}{\textwidth}
        \includegraphics[width=0.45\linewidth]{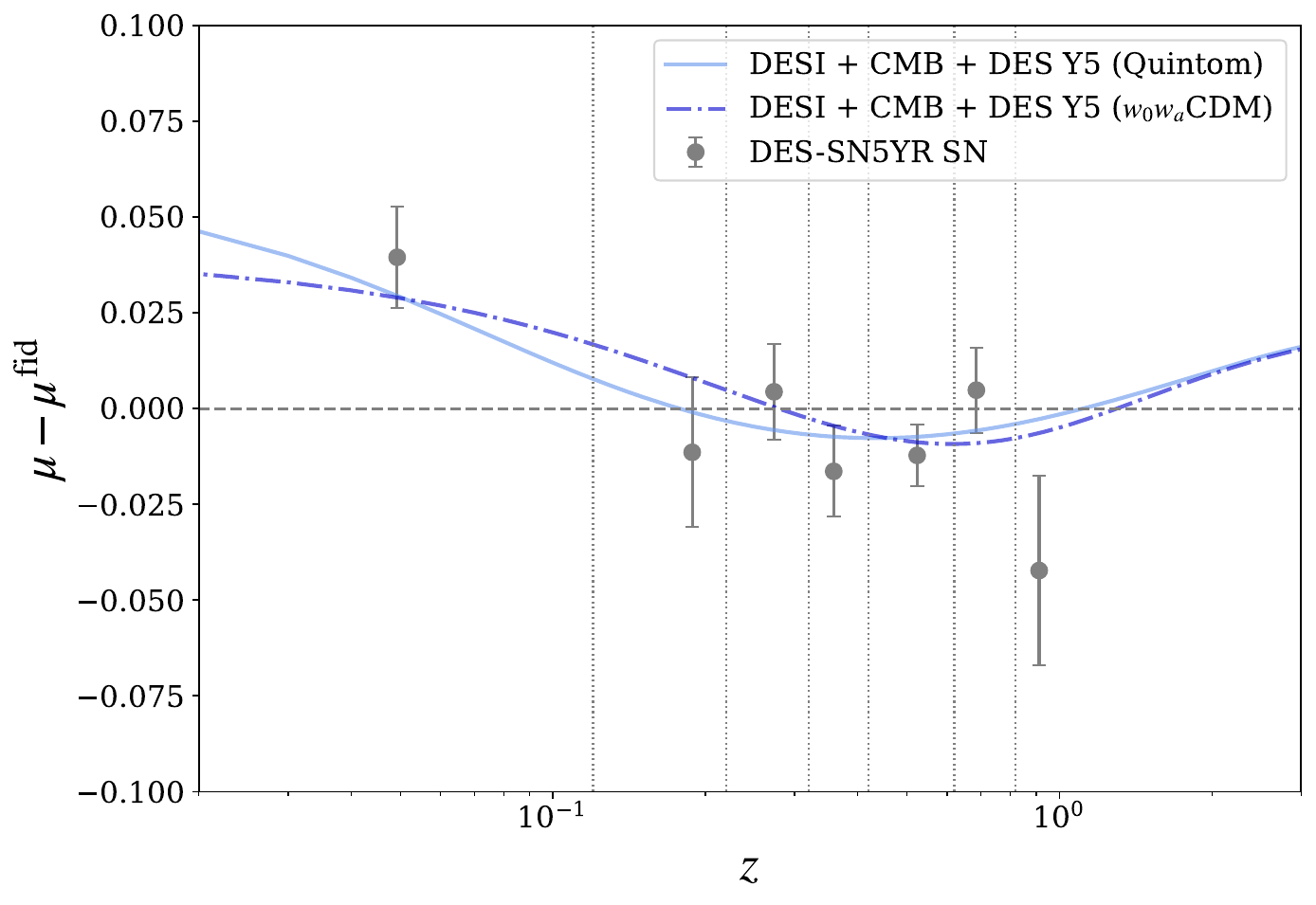}
    \end{subfigure}
    \vspace{0.5cm} 
    \begin{subfigure}{\textwidth}
        \includegraphics[width=0.45\linewidth]{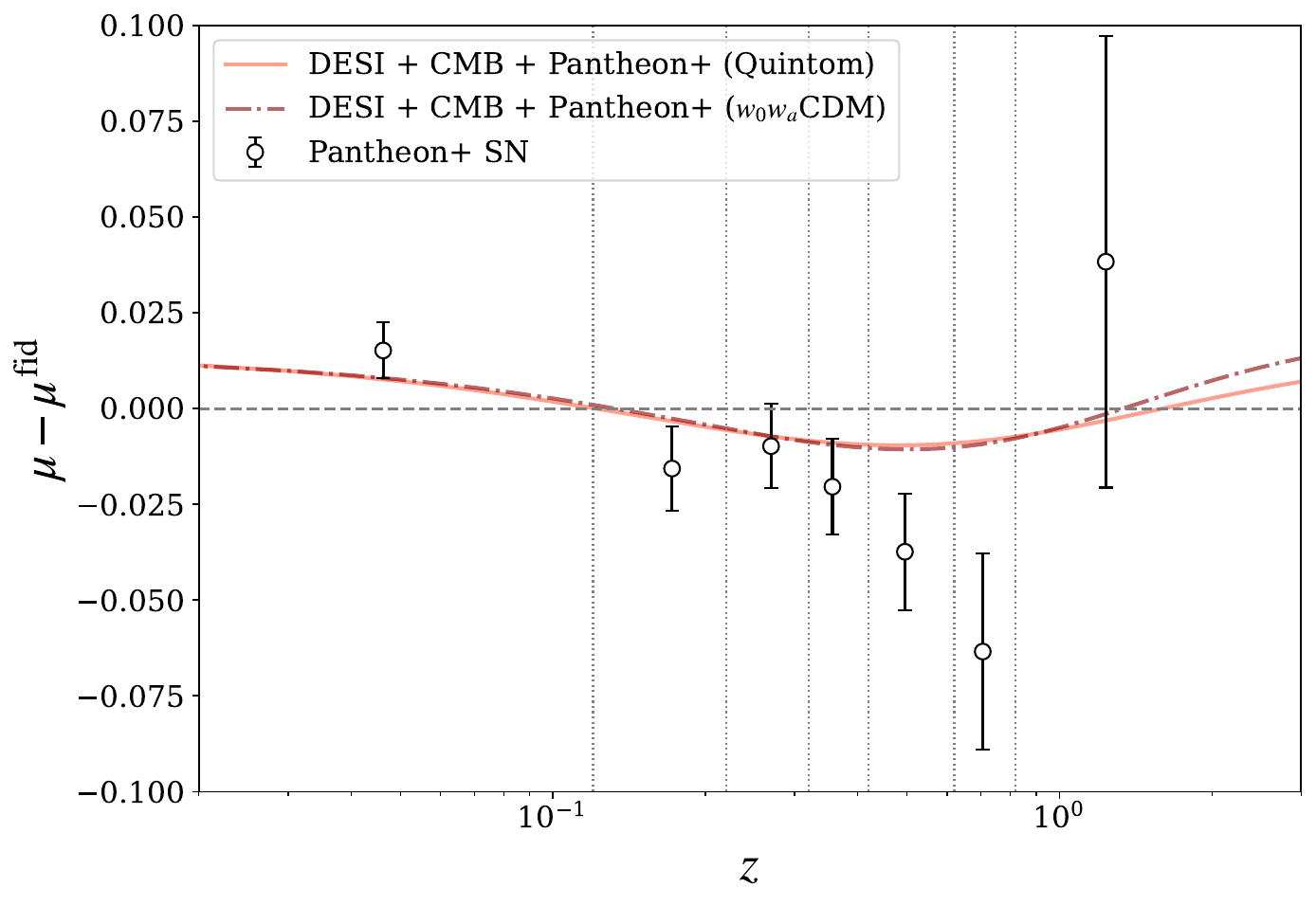}
    \end{subfigure}
    \caption{Plot of the distance modulus residuals as a function of redshift, assuming \textit{Planck} 2018 $\Lambda$CDM fiducial cosmology, using the best-fit values for each model (quintom in solid or $w_0w_a$CDM in dash-dot) and dataset (top: DESY5, bottom: Pantheon+). We plot the binned residuals of the SNe1a data, following the methodology highlighted in Sect. IV C of \citet{2025arXiv250314738D}, where the vertical grey lines denote the bin edges.}
    \label{fig:sn_residuals} 
\end{figure}

In Fig.~\ref{fig:sn_residuals}, we additionally show the residuals of the Type Ia supernova distance modulus, $\mu-\mu^{\rm fid}$, relative to the fiducial \textit{Planck} $\Lambda$CDM cosmology, following the binning and normalisation procedure adopted in \citet{2025arXiv250314738D}. Once again, the quintom model is able to reproduce the observed variation between the low- and high-redshift supernova measurements, particularly for the DESY5 dataset. Together with the systematically lower BAO distance measurements, they drive the preference for a dynamical dark energy model.

To quantitatively compare the quintom and $w_0w_a$CDM cosmologies while accounting for the differing number of free parameters, we evaluate the Bayesian evidence, $\log\mathcal{Z}$, defined as the probability of obtaining the observed data under a given model hypothesis. The evidence is computed by marginalising the likelihood $p(d|\Theta,M)$ over the parameter space $\Theta$, weighted by the prior distribution $p(\Theta|M)$,
\begin{equation}    
\mathcal{Z}=p(d|M) = \int p(d|\Theta,M)\,p(\Theta|M)\,d\Theta\,.
\end{equation}
The relative preference between two models can then be quantified through the Bayes factor,
\begin{equation}    
B_{1-2}=\frac{\mathcal{Z}_1}{\mathcal{Z}_2}\,,
\end{equation}
which compares the probability of the data under the two competing hypotheses \citep{2015DSP....47...50K}. The resulting log-evidence values and Bayes factors are summarised in Table~\ref{tab:logz}. According to the Jeffreys scale \citep{jeffreys1939theory}, we find that $B_{{\rm quint}-w_0w_a}\approx3$ for both dataset combinations, corresponding to a moderate preference for the quintom cosmology over the phenomenological $w_0w_a$CDM parametrisation, despite the larger parameter volume associated with the additional scalar field degrees of freedom.

\begin{table}
    \centering
    \fontsize{9}{13}\selectfont
    \begin{tabular}{|c c c|}
    \hline
     & 
    \makecell{DESI+CMB \\ +DESY5} &
    \makecell{DESI+CMB \\ +Pantheon+} \\
    \hline
    Quintom & $-1159.77$ & $-1043.89$ \\
    $w_0w_a$CDM & $-1163.60$ & $-1047.15$ \\
    \hline
    $B_{{\rm quint}-w_0w_a}$ & 3.83&3.26 \\
    \hline
    \end{tabular}
    \caption{Log-evidence values and Bayes' factor for the models under different dataset combinations.}
    \label{tab:logz}
\end{table}

\subsection{Parameter Degeneracies in a Quintom Model}
\begin{figure*}
    \centering
    \includegraphics[width=0.85\linewidth]{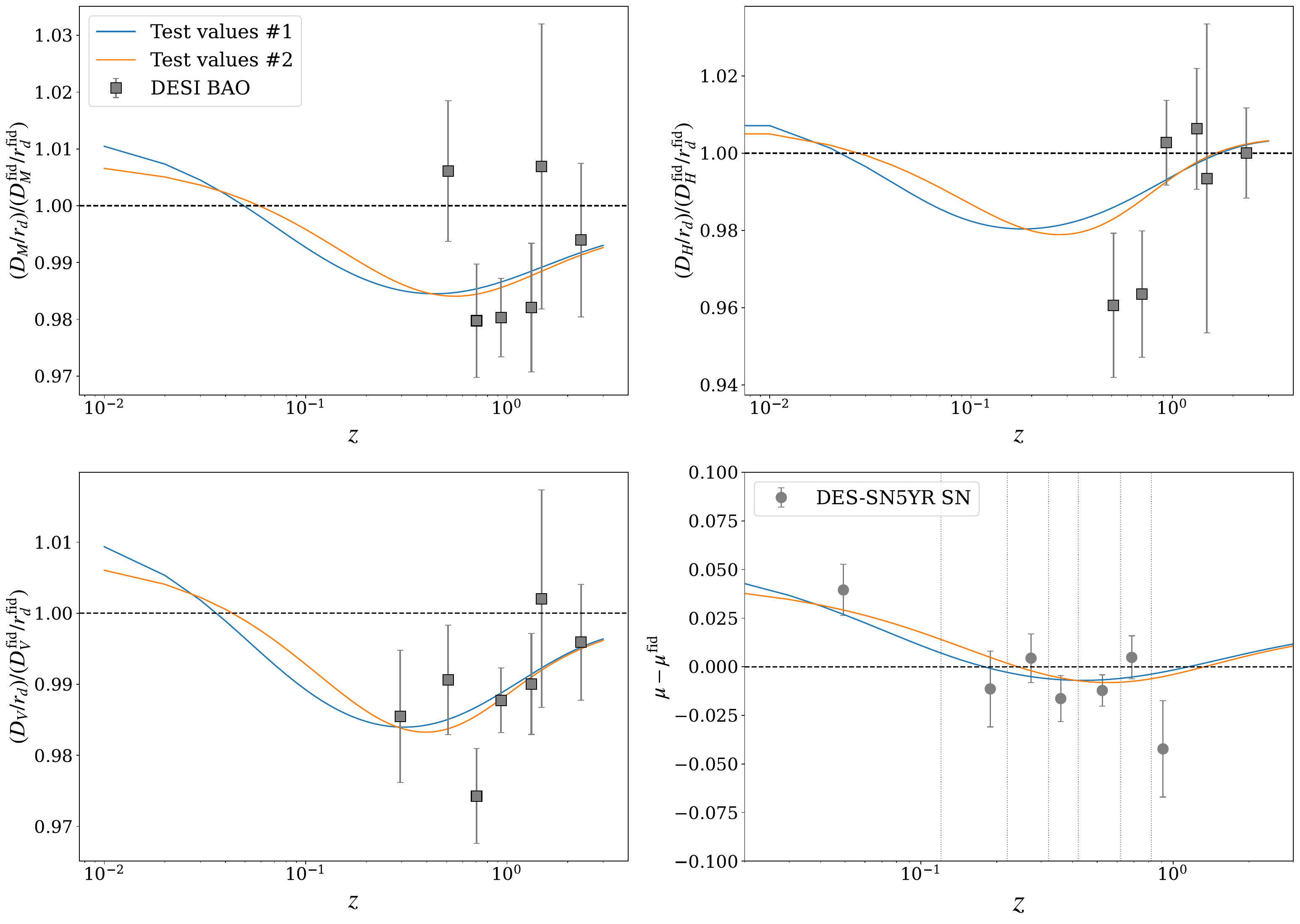}
    \caption{Plot of (clockwise from top left): $(D_M /r_d)/(D^{\rm{fid}}_M /r^{\rm{fid}}_d)$, $(D_H /r_d)/(D^{\rm{fid}}_H /r^{\rm{fid}}_d)$, $\mu-\mu^{\rm fid}$ and $(D_V /r_d)/(D^{\rm{fid}}_V /r^{\rm{fid}}_d)$ for the two parameter sets with similar log-likelihood values (blue and orange). We plot the DESI BAO data points in grey squares, and the binned DESY5 SNe1a data in grey circles. }
    \label{fig:distance_degen}
\end{figure*}

\begin{figure}
    \centering
    \includegraphics[width=\linewidth]{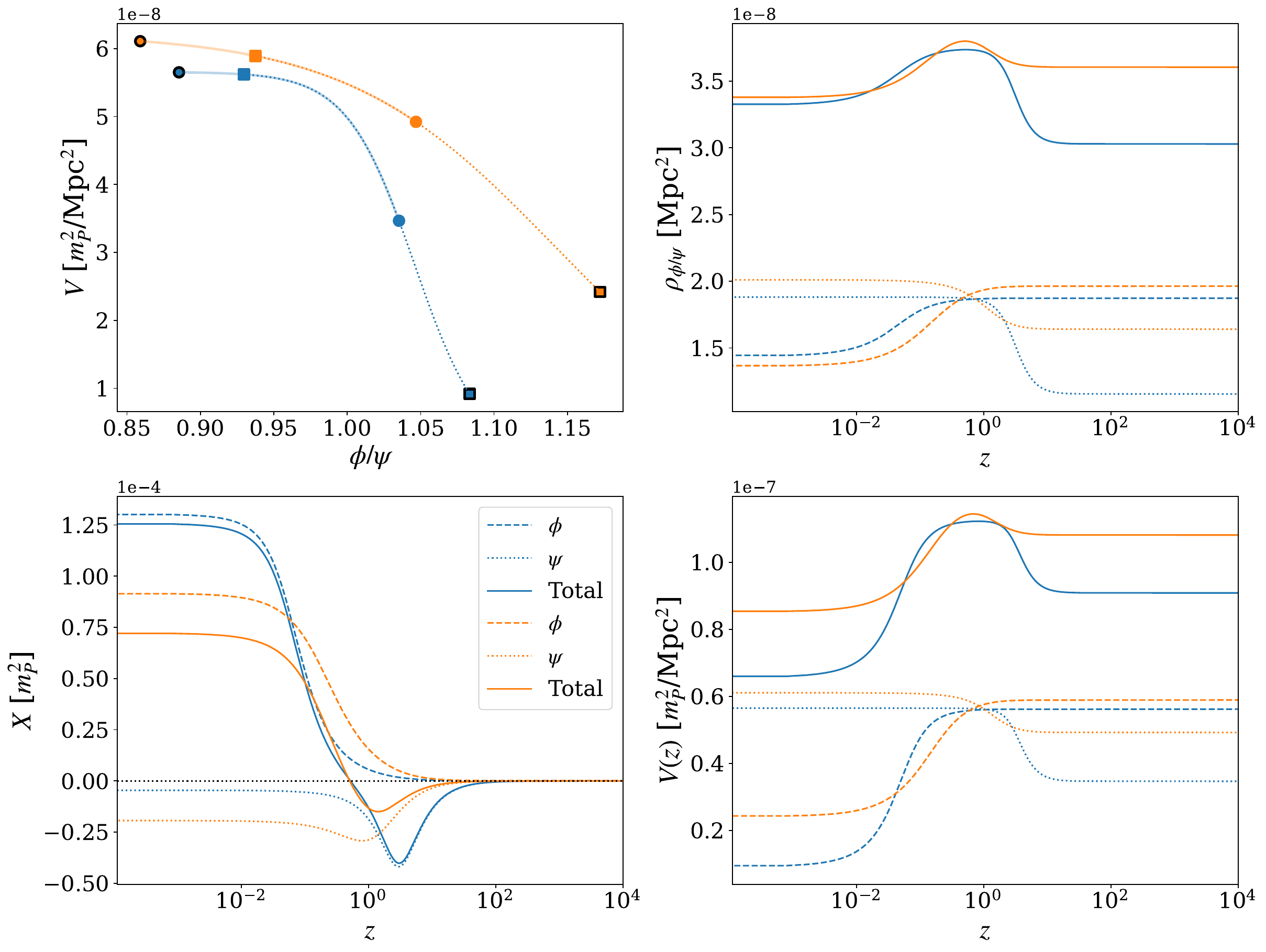}
    \caption{Top left: Plot of the potential $V(\phi/\psi)$ for the two parameter sets (blue and orange). We also plot the starting (in squares) and ending (in circles) points of the $\phi$ (points with no border) and $\psi$ (points with black edges) field as they evolve along the potential. Clockwise from top right: Plot of the evolutions of the scalar field energy densities $\rho_{\phi/\psi}(z)$, the potential $V(z)$, and their speeds $X(z)$. We plot the total quantities in solid lines, those of the $\phi$ field in dashed lines, and those of the $\psi$ field in dotted lines.}
    \label{fig:v_degen}
\end{figure}

As highlighted at the beginning of this section, the quintom parameter space exhibits a number of interesting degeneracies, most notably the characteristic ``boomerang-shaped'' posterior distributions appearing in the $s-\phi_{\rm ini}$ and $s-V_0$ parameter planes. Similar behaviour was previously identified in \citet{2025MNRAS.544.3142G}, where we demonstrated that different combinations of $s$ and $\phi_{\rm ini}$ lying along an approximate degeneracy direction can yield nearly identical values of the effective equation-of-state parameter $w_0$.

To investigate this further, we select two representative samples from the DESI+CMB+DESY5 MCMC chain that possess nearly identical likelihoods (differing by less than $0.01\%$) while simultaneously maximising the Euclidean separation in the three-dimensional parameter space spanned by $(s,V_0,\phi_{\rm ini})$. The corresponding parameter values are listed in Table~\ref{tab:degen_vals}, while the resulting BAO distance measures and SNe1a distance moduli are shown in Fig.~\ref{fig:distance_degen}. We see that despite different values of $V_0, s$ and $\phi_{\rm ini}$, they produce almost identical curves that fit the data equally well. The physical origin of this degeneracy becomes clearer when examining the scalar field dynamics directly. In Fig.~\ref{fig:v_degen}, we plot both the shape of the potential $V(\phi)$ and the evolutions of their speeds and energy densities. 

\begin{table}
    \centering
    \fontsize{9}{13}\selectfont
\begin{tabular}{|c c c|}
\hline
    \textbf{Parameter} & 
        Parameter Set 1 &
        Parameter Set 2\\
         \hline
         $\omega_{\rm c}$& 0.1183 & 0.1178 \\
         $\omega_{\rm b}$& 0.02246 & 0.02251\\
         $n_{\rm s}$&0.971&0.969\\
         $\ln{(10^{10}A_{\rm s})}$& 3.091& 3.096\\
         $\tau$& 0.080&0.084\\
         $V_0\times 10^{8}\,[m_{\rm P}^4]$& 2.828 & 3.109\\
         $s\,[m_{\rm P}^{-1}]$&21.83&7.120\\
         $\phi_{\rm ini}\,[m_{\rm P}]$& 0.930&0.937\\
         $\psi_{\rm ini}\,[m_{\rm P}]$&1.035&1.047\\
         \hline
         $\chi^2$&$-1131.44$&  $-1131.52$\\
         \hline
    \end{tabular}
    \caption{The two sets of parameter values we extracted from our chain to investigate the degeneracies between the quintom model parameters.}
    \label{tab:degen_vals}
\end{table}

Increasing the steepness parameter $s$ causes the fields to accelerate more rapidly along the potential. However, from the approximate relation $w_{\rm quintom}\approx -1+2\epsilon\,,$
the crossing behaviour is determined mainly by the ratio between the effective kinetic contribution,
$\dot{\phi}^2-\dot{\psi}^2\,,$ and the total potential energy. As a result, different combinations of $\dot{\phi}$, $\dot{\psi}$, $V_0$, and $s$ can lead to very similar evolutions of $\epsilon$, depending on both the slope of the potential and the initial field positions.

The dynamics of this mechanism are much more sensitive to $\phi_{\rm ini}$ than $\psi_{\rm ini}$, as the phantom dominated phase takes place over a longer duration, allowing the $\psi$ field more time to evolve. The expansion history from Eq. \eqref{eq:friedmann} itself cannot break this degeneracy either: we can obtain the same quantity for $\rho_{\phi}+\rho_{\psi}=X_{\phi}-X_{\psi}+V(\phi)+V(\psi)$, whereby a larger $X_{\phi}-X_{\psi}$ (driven by a larger value of $s$) and is compensated by a smaller $V_0$, which is precisely the behaviour being captured in this example. This degeneracy therefore appears to be an intrinsic feature of the scalar field potentials, rather than a phenomenological one. 

\section{Conclusion}\label{sec:conclusions}

We have extended our previous work presented in \citet{2025MNRAS.544.3142G} by investigating the perturbation dynamics of a two-field quintom cosmology. To do so, we derived the relevant perturbation equations and modified the Boltzmann solver \texttt{CLASS} to incorporate a quintessence field and a phantom field evolving along a hyperbolic tangent potential, and studied the resulting evolution of matter and scalar field perturbations in the framework of a phantom-to-quintessence transition.

We find that modifications to the expansion history are the dominant source of changes to structure growth for a fixed cosmological background. At early times, the total scalar field energy density is lower than the corresponding $\Lambda$CDM value, leading to a reduced expansion rate and consequently enhanced matter clustering. During the phantom-dominated phase, however, the combined dark energy density $\rho_\phi+\rho_\psi$ increases and eventually exceeds $\rho_\Lambda$, producing a temporary enhancement in $H(z)$ and a corresponding suppression in the growth of $\delta\rho_{\rm m}$. At later times, once the quintessence field becomes dominant, $\rho_\phi+\rho_\psi$ decreases and the expansion rate decreases again.

We also examined the behaviour of the scalar field perturbations and found them to be strongly scale-dependent. Intermediate scales exhibit the largest density and pressure perturbations, since large scale modes remain super-horizon for longer and therefore have less time to evolve, while small scale modes enter the horizon earlier and are suppressed by Hubble friction. Nevertheless, the scalar field perturbations remain significantly smaller in amplitude than the matter perturbations, and therefore have only a limited direct impact on structure formation. Correspondingly, when studying the gravitational potentials and the resulting LSS and CMB observables, we find that deviations from $\Lambda$CDM are driven primarily by changes in $\delta_{\rm m}$. In particular, the quintom cosmology produces a $1$--$2\%$ suppression in the linear matter power spectrum across all scales, together with an enhancement in the large-scale CMB temperature anisotropies through the late-time Integrated Sachs-Wolfe effect.

We subsequently performed a Bayesian inference analysis of both the quintom and $w_0w_a$CDM models, for a DESI+CMB+DESY5 and a DESI+CMB+Pantheon+ dataset. We recover similar best-fit values for the standard cosmological parameters in both cases, while the best-fit quintom parameters naturally reproduce the desired phantom-to-quintessence transition. We also find good agreement between the best-fit BAO distance measures and supernova distance moduli predicted by the two models for both dataset combinations. When conducting model comparison by calculating the log-evidence of each model and dataset combination, we find that a quintom model is moderately favoured over a $w_0w_a$CDM parametrisation, despite its larger prior volume.

Finally, we investigated the parameter degeneracies between $V_0$, $s$, and $\phi_{\rm ini}$, finding that they arise primarily from the chosen expression of the hyperbolic tangent potential itself. Larger values of $V_0$ can be compensated by shallower gradients, leading to similar relative differences between the velocities of the quintessence and phantom fields, and therefore similar evolutions of $w(z)$ and the cosmological distances. Since these combinations produce nearly identical background expansion histories, the degeneracy cannot be efficiently broken using background observables alone.

Overall, the quintom model reproduces many of the same phenomenological trends as a $w_0w_a$CDM parametrisation exhibiting phantom-to-quintessence crossing, making it a compelling candidate as a theoretically motivated explanation of a $w_0w_a$CDM parametrisation. It would also be valuable to include LSS data such as weak gravitational lensing and galaxy clustering, which directly constrain the amplitude of matter clustering at late times. This would require accurate nonlinear modelling for these scalar field models, for example with \texttt{ReACT} \citep{2020MNRAS.498.4650B}, which uses a halo-model reaction approach to predict the nonlinear power spectrum boost arising from beyond $\Lambda$CDM physics.

We are entering an era of increasingly precise cosmological observations, with ongoing galaxy surveys such as \textit{Euclid} \citep{2025A&A...697A...1E}, LSST \citep{2019ApJ...873..111I}, and CMB data from the upcoming Simons Observatory \citep{2024ApJS..274...33G}, and LiteBIRD \citep{2020SPIE11443E..2FH} missions. In this context, physically motivated models such as the quintom scenario explored in this work provide an important framework for interpreting potential deviations from $\Lambda$CDM, allowing us to probe the underlying physics of dark energy and perhaps answer one of the most pertinent questions in cosmology today.

\section*{Acknowledgements}
LWKG thanks the University of Edinburgh School of Physics and Astronomy for a postdoctoral Fellowship, while ANT thanks the UK Science and Technology Funding Council (STFC) for support. For the purpose of open access, the authors have applied a Creative Commons Attribution (CC BY) licence to any Author Accepted Manuscript version arising from this submission.

\section*{Data Availability}
The data underlying this article will be shared on reasonable request to the corresponding author.




\bibliographystyle{mnras}
\bibliography{biblio} 




\bsp	
\label{lastpage}
\end{document}